\documentclass[aps,prb,twocolumn,showpacs,10pt,superscriptaddress,floatfix,longbibliography]{revtex4-2}

\usepackage{newtxtext}
\usepackage{newtxmath}
\usepackage{graphicx}
\usepackage{subfigure}
\usepackage{amsmath}
\usepackage{amsfonts}
\usepackage{mathrsfs}
\usepackage{float}
\usepackage{booktabs}
\usepackage{bbm}
\usepackage{xcolor}
\usepackage[colorlinks=true]{hyperref}
\usepackage{ulem}
\usepackage[T1]{fontenc}
\usepackage{svg}
\usepackage{appendix}
\usepackage{comment}

\begin{document}
\title{Berry curvature effects of chiral superconducting rhombohedral graphene}
\author{Jian-Hua Zeng}
\affiliation{School of Physics, Sun Yat-sen University, Guangzhou 510275, China}

\author{Zhi Wang}
\email{wangzh356@mail.sysu.edu.cn}

\affiliation{School of Physics, Sun Yat-sen University, Guangzhou 510275, China}
\affiliation{Guangdong Provincial Key Laboratory of Magnetoelectric Physics and Devices, Sun Yat-sen University, Guangzhou 510275, China}

\author{Qian Niu}
\affiliation{Department of Physics, University of Science and Technology of China, Anhui 230026, China}

\begin{abstract}
We study the Berry curvature effects of the Bogoliubov quasiparticles in chiral superconducting rhombohedral graphene. Using a two-band Bogoliubov-de Gennes Hamiltonian to describe the superconducting quasiparticles, we calculate the momentum-space Berry curvature, orbital magnetic moment, and anomalous thermal Hall, spin Nernst, and orbital Nernst transport of chiral $p$-wave superconducting states. We investigate the impact of normal-state band warping in rhombohedral graphene, which can generate Bogoliubov Fermi surfaces for quasiparticle excitations. We find that Bogoliubov Fermi surfaces qualitatively modify the anomalous transport responses, inducing a deviation of the thermal Hall conductivity from the quantized value and strongly enhancing the spin and orbital Nernst responses. 
\end{abstract}
\maketitle

\section{Introduction}
Chiral superconductivity is an unconventional superconducting state with spontaneous time reversal symmetry breaking and finite-angular-momentum Cooper pairing~\cite{RPP2016Kallin_ChiralSC}. Such a state has attracted sustained interest because it provides a possible route to topological superconductivity and Majorana zero modes~\cite{PRB2000Read_Majorana,JPSJ2012Tanaka_TSC,JPSJ2016Sato_Majorana,RPP2017Sato_Majorana,PTEP2024Tanaka_Majorana}, with potential applications in quantum computation~\cite{RMP2008Nayak_QuanCom}. The rapid development of two-dimensional materials has opened up new opportunities to search for chiral superconductivity in highly tunable platforms. Superconductivity has by now been observed in a broad range of two-dimensional systems, including twisted multilayer graphene~\cite{Nat2018Cao_twist,Sci2019Matthew_twist,Nat2019Lu_twist,Nat2020Arora_twist,Nat2021Cao_twist,Nat2021Oh_twist,Nat2021Park_twist,Sci2021Hao_twist,Sci2022Zhang_twist}, moiréless multilayer graphene~\cite{Nat2021Zhou_moireless,Sci2022Zhou_moireless,Nat2023Zhang_moireless,Nat2024Li_moireless,Nat2025Han_RHG,Nat2025Choi_moireless,NatMat2025Yang_moireless}, as well as twisted bilayer ${\mathrm{WSe}}_{2}$~\cite{Nat2025Xia_WSe2,Nat2025Guo_WSe2}, significantly expanding the material landscape for investigating unconventional superconductivity.

Among these systems, rhombohedral multilayer graphene is particularly attractive because its low-energy bands can be electrically tuned and flattened by displacement fields~\cite{PRB2009Koshino_RHG,PRB2010Zhang_RHG}, strongly enhancing interaction effects without relying on moiré superlattices. It has thus emerged as a promising setting for a variety of correlated phenomena~\cite{PRL2011Zhang_RHG,NatPhy2011Lui_RHG,NatPhy2011Bao_RHG,NanoLett2013Zou_RHG,Nat2020Shi_RHG,Nat2021Zhou_QM,NatNano2024Han_RHG,NatNano2024Liu_RHG,Sci2024Han_RHG,Sci2024Sha_RHG}.
Recently, superconductivity was discovered in rhombohedral tetralayer graphene, with several unusual features such as magnetic hysteresis, robustness against large in-plane magnetic fields, and zero-field anomalous Hall response in the normal state~\cite{Nat2025Han_RHG}. Together, these features suggest a chiral superconducting state. This discovery has stimulated many theoretical studies of the pairing mechanism~\cite{PRB2025Chou_theory,PRB2025Yang_theory,PRB2025Parra_theory,NC2025Geier_theory,PRB2026Jahin_theory,PRL2026Yoon_RHG,CPL2026Qin_theory}, and chiral $p$-wave superconductivity, with pairing within a single spin- and valley-polarized sector, has been widely discussed as a natural candidate.

Geometric effects in superconductors have attracted growing attention in recent years~\cite{NC2015Cvetkovic_TH,PRB2015Murray_SCBC,NatPhy2019Qin_SCGeo,PRL2021Wang_SCBC,PRB2021Villegas_SCGeo,PRR2022Kitamura_SCGeo,PRB2022Simon_SCBC,Nat2023Tian_SCGeo,PRL2023Jiang_SCGeo,PRB2023Zhou_SCBC,PRB2023Zhang_SCBC,PRL2024Zhang_SCGeo,PRL2024Chen_SCGeo,PRB2024Daido_SCGeo,Arxiv2024Liao_SCBC,PRL2025Shavit_SCGeo,PRL2025Porlles_SCGeo,PRB2025Simon_SCGeo,PRB2025Liao_SCBC,Arxiv2025matsushita_SCBC,PRL2026Hsieh_SCBC,PRB2026Liao_SCBC}. Among them, the Berry curvature of superconducting quasiparticles has received increasing attention, as it can lead to an anomalous Hall response even in a topologically trivial state. Chiral superconductors are well suited for studying these effects because they break time reversal symmetry and can therefore support finite Berry curvature and related transverse responses. In this context, rhombohedral graphene is a particularly appealing platform. Chirally stacked few-layer graphene has long been discussed in connection with spontaneous quantum Hall physics~\cite{PRL2011Zhang_RHG}, and recent experiments have reported signatures of chiral superconductivity emerging next to spin- and valley-polarized states whose normal state already shows anomalous Hall response~\cite{Nat2025Han_RHG}. This makes it natural to examine the Berry-curvature-related transport of superconducting quasiparticles in this system. Besides Berry curvature, another closely related quantity is the orbital magnetic moment of Bogoliubov quasiparticles~\cite{PRB2026Zeng_omm}, which is also relevant in this system. Experiments on rhombohedral tetralayer graphene have revealed unusual magnetic properties~\cite{Nat2025Han_RHG}, and recent work has further highlighted rhombohedral graphene as a promising platform for studying orbital magnetic response in chiral superconductors~\cite{Arxiv2026Zhu_SCOM}.

Another important feature of intravalley pairing in rhombohedral graphene is the possible appearance of Bogoliubov Fermi surfaces~\cite{PRB2025Yang_theory,PRL2026Yoon_RHG}. 
When the normal-state dispersion satisfies $\xi_{\bf k}=\xi_{-\bf k}$, chiral pairing generally opens a full quasiparticle gap~\cite{NC2025Geier_theory}. 
Here $\mathbf{k}$ is measured from the $K$ point in the Brillouin zone.
However, the realistic band structure of rhombohedral graphene contains trigonal warping, which makes the intravalley dispersion asymmetric, $\xi_{\bf k}\neq \xi_{-\bf k}$. 
This asymmetry shifts the Bogoliubov bands in energy, so that the quasiparticle spectrum can remain gapless even when the pairing amplitude is finite. 
Such Bogoliubov Fermi surfaces modify the occupied quasiparticle states in momentum space and can therefore strongly affect Berry-curvature-induced transport.

In this work, we study the Berry curvature, orbital magnetic moment, and thermal transport of superconducting quasiparticles in chiral superconducting rhombohedral graphene. 
We describe the superconducting quasiparticles by a two-band BdG Hamiltonian. 
We then compare two physical situations. 
The first is a fully gapped chiral superconducting state, for which we calculate the quasiparticle Berry curvature, thermal Hall conductivity, and spin Nernst coefficient. 
The second is a gapless chiral superconducting state with Bogoliubov Fermi surfaces, for which we examine how the Bogoliubov Fermi surfaces affect the quasiparticle Berry curvature, orbital magnetic moment, thermal Hall conductivity, spin Nernst coefficient, and orbital Nernst coefficient. 
We show that Bogoliubov Fermi surfaces lead to a nonquantized low-temperature thermal Hall response and strongly enhance the low-temperature spin and orbital Nernst responses. 
We also study the electron density dependence of these transport coefficients and show that their magnitudes and signs are sensitive to the presence, size, and momentum-space location of the Bogoliubov Fermi surfaces.

The rest of this paper is organized as follows. 
In Sec.~\ref{sec:model_method}, we introduce the two-band BdG Hamiltonian for superconducting quasiparticles in rhombohedral graphene and define the Berry curvature, orbital magnetic moment, thermal Hall conductivity, spin Nernst coefficient, and orbital Nernst coefficient. 
In Sec.~\ref{sec:no_BFS}, we discuss the Berry curvature and thermal transport in fully gapped superconducting states. 
In Sec.~\ref{sec:with_BFS}, we study the Berry curvature, orbital magnetic moment, and thermal transport in gapless superconducting states with Bogoliubov Fermi surfaces. 
Finally, Sec.~\ref{sec:conclusion} summarizes the main results.

\section{Model and method}
\label{sec:model_method}
We formulate the low-energy superconducting quasiparticles in rhombohedral graphene using a two-band BdG Hamiltonian. 
The normal-state dispersion is projected onto the lowest conduction band, and the superconducting state is described within a single spin and valley sector. 
We take this active sector to be the spin-up sector near the $K$ valley and omit the spin index below. 
Using the Nambu spinor $\Psi_{\mathbf{k}}=(c_{\mathbf{k}},c_{-\mathbf{k}}^\dagger)^T$, the BdG Hamiltonian is written as
\begin{equation}
    \hat{H}_{\mathbf{k}}=
    \begin{pmatrix}
        \xi_{\mathbf{k}} & \Delta_{\mathbf{k}} \\
        \Delta_{\mathbf{k}}^{*} & -\xi_{-\mathbf{k}}
    \end{pmatrix},
    \label{eq:generic_BdG}
\end{equation}
where $\mathbf{k}$ is measured from the $K$ point, $\xi_{\mathbf{k}}=\varepsilon_{\mathbf{k}}-\mu$ is the normal-state energy measured from the chemical potential $\mu$, and $\Delta_{\mathbf{k}}$ is the superconducting gap function with the chiral $p$-wave pairing~\cite{NC2025Geier_theory,PRL2026Yoon_RHG}. 
The concrete forms of $\xi_{\mathbf{k}}$ and $\Delta_{\mathbf{k}}$ for the models studied in this work are given in Appendices~\ref{app:model_without_BFS} and \ref{app:model_with_BFS}.

We separate the normal-state dispersion into symmetric and antisymmetric parts,
\begin{equation}
    \xi_s(\mathbf{k})=\frac{\xi_{\mathbf{k}}+\xi_{-\mathbf{k}}}{2},
    \qquad
    \xi_a(\mathbf{k})=\frac{\xi_{\mathbf{k}}-\xi_{-\mathbf{k}}}{2}.
\end{equation}
Then Eq.~\eqref{eq:generic_BdG} becomes
\begin{equation}
    \hat{H}_{\mathbf{k}}
    =
    \xi_a(\mathbf{k})\tau_0
    +
    \xi_s(\mathbf{k})\tau_z
    +
    \Delta^{a}_{\mathbf{k}}\tau_x
    -
    \Delta^{b}_{\mathbf{k}}\tau_y ,
    \label{eq:BdG_Pauli}
\end{equation}
with $\Delta^{a}_{\mathbf{k}}=\mathrm{Re}\Delta_{\mathbf{k}}$ and $\Delta^{b}_{\mathbf{k}}=\mathrm{Im}\Delta_{\mathbf{k}}$, and the two quasiparticle energies are
\begin{equation}
    E_{\pm,\mathbf{k}}
    =
    \xi_a(\mathbf{k})
    \pm
    \sqrt{
    \xi_s^2(\mathbf{k})+|\Delta_{\mathbf{k}}|^2
    } .
    \label{eq:BdG_spectrum}
\end{equation}
If $\xi_{\mathbf{k}}=\xi_{-\mathbf{k}}$, one has $\xi_a(\mathbf{k})=0$. In this case, the quasiparticle spectrum in Eq.~\eqref{eq:BdG_spectrum} is generally fully gapped for the model considered in Sec.~\ref{sec:no_BFS}, where trigonal warping is neglected. By contrast, trigonal warping makes $\xi_{\mathbf{k}}\neq\xi_{-\mathbf{k}}$ in the intravalley pairing state. 
The corresponding antisymmetric dispersion $\xi_a(\mathbf{k})$ can shift the Bogoliubov branches and, for suitable parameters, generate zero-energy contours, namely Bogoliubov Fermi surfaces.
This provides the basis for comparing the model without Bogoliubov Fermi surfaces in Sec.~\ref{sec:no_BFS} and the model with Bogoliubov Fermi surfaces in Sec.~\ref{sec:with_BFS}.

We now define the geometric quantities and transport coefficients calculated below. 
The Berry curvature of the $n$th Bogoliubov band is~\cite{PRL2021Wang_SCBC,PRB2023Zhang_SCBC,PRB2023Zhou_SCBC,Arxiv2024Liao_SCBC}
\begin{equation}
\mathbf{\Omega}_{n}(\mathbf{k})=-\sum_{n^{\prime}\neq n}\mathrm{Im}\left[\frac{\langle\phi_{n}|\partial_{\mathbf{k}}\hat{H}_{\mathbf{k}}|\phi_{n^{\prime}}\rangle\times\langle\phi_{n^{\prime}}|\partial_{\mathbf{k}}\hat{H}_{\mathbf{k}}|\phi_{n}\rangle}{\left(E_{n^{\prime},\mathbf{k}}-E_{n,\mathbf{k}}\right)^{2}}\right],\label{eq:BC}
\end{equation}
where $E_{n,\mathbf{k}}$ and $|\phi_{n,\mathbf{k}}\rangle$ are the eigenenergy and eigenstate of $\hat{H}_{\mathbf{k}}$.

The orbital magnetic moment of the Bogoliubov quasiparticle is calculated from~\cite{PRB2026Zeng_omm}
\begin{equation}
\mathbf{m}_{n}(\mathbf{k})=\frac{e}{2\hbar}\sum_{n^{\prime}\neq n}\mathrm{Im}\left[\frac{\langle\phi_{n}|\partial_{\mathbf{k}}\hat{H}_{\mathbf{k}}|\phi_{n^{\prime}}\rangle\times\langle\phi_{n^{\prime}}|\tau_{z}\partial_{\mathbf{k}}\hat{H}_{\mathbf{k}}^{d}|\phi_{n}\rangle}{E_{n^{\prime},\mathbf{k}}-E_{n,\mathbf{k}}}\right],\label{eq:OMM}
\end{equation}
where $-e$ is the electron charge, and $\hat{H}_{\mathbf{k}}^{d}$ denotes the diagonal part of the BdG Hamiltonian,
\begin{equation}
    \hat{H}_{\mathbf{k}}^{d}
    =
    \begin{pmatrix}
        \xi_{\mathbf{k}} & 0 \\
        0 & -\xi_{-\mathbf{k}}
    \end{pmatrix}
    =
    \xi_a(\mathbf{k})\tau_0+\xi_s(\mathbf{k})\tau_z .
\end{equation}
Therefore,
\begin{equation}
    \tau_z\partial_{\mathbf{k}}\hat{H}_{\mathbf{k}}^{d}
    =
    \partial_{\mathbf{k}}\xi_s(\mathbf{k})\tau_0
    +
    \partial_{\mathbf{k}}\xi_a(\mathbf{k})\tau_z .
    \label{eq:tauz_dH}
\end{equation}
In the symmetric case $\xi_{\mathbf{k}}=\xi_{-\mathbf{k}}$, the second term in Eq.~\eqref{eq:tauz_dH} vanishes, and the remaining identity term gives no interband matrix element because of the orthogonality of the Bogoliubov eigenstates. 
Thus, within the two-band Hamiltonian, a finite quasiparticle orbital magnetic moment requires an asymmetric normal-state dispersion.

The Berry curvature further gives rise to anomalous Hall-type transport of superconducting quasiparticles. We first consider the thermal Hall effect, and the thermal Hall conductivity is written as~\cite{NC2015Cvetkovic_TH,PRR2020Teng_THE,PRL2021Wang_SCBC,PRB2023Zhang_SCBC,PRB2023Zhou_SCBC}
\begin{equation}
\kappa_{xy}=\frac{1}{\hbar T}\sum_{n}\int[d\mathbf{k}]\,\Omega^{z}_{n}(\mathbf{k})\int_{E_{n,\mathbf{k}}}^{\infty}d\epsilon\,\epsilon^{2}f'(\epsilon,T),\label{eq:TH}
\end{equation}
where $f(E,T)=1/(e^{E/k_{{\rm B}}T}+1)$ is the Fermi distribution function, $f'(\epsilon,T)\equiv\partial f(\epsilon,T)/\partial\epsilon$, and $\int[d\mathbf{k}]\equiv(1/2\pi)^{2}\int d^{2}k.$ 
In the present two-dimensional system, only the out-of-plane component of the quasiparticle Berry curvature is nonzero. 

We also calculate the spin Nernst coefficient. The spin operator in particle-hole space is $\mathbf{S}=\mathbf{s}\otimes\tau_{z}$, where $\mathbf{s}$ represents the electron spin. For the quarter-metal normal state considered here, we take $s_{x}=s_{y}=0$ and $s_{z}=\hbar/2$, so
that $S^{z}=\hbar\tau_{z}/2$. The spin Nernst coefficient is then written
as~\cite{PRB2021Xiao_ConservedCurrent,Arxiv2024Liao_SCBC,PRB2026Liao_SCBC} 
\begin{equation}
\alpha_{\mathrm{SN}}=\frac{1}{\hbar}\sum_{n}\int[d\mathbf{k}]\,\frac{\partial g(E_{n,\mathbf{k}},T)}{\partial T}\,\left\langle S^{z}\right\rangle \,\Omega^z_{n}(\mathbf{k}),\label{eq:SN}
\end{equation}
where $g(E,T)=k_{{\rm B}}T\ln\!\left[1-f(E,T)\right]$ is the quasiparticle grand-potential function, and $\left\langle ...\right\rangle $ denotes the expectation value taken with respect to the Bogoliubov eigenstates.

Finally, when the quasiparticle orbital magnetic moment is finite, we evaluate the orbital Nernst coefficient as~\cite{PRB2026Zeng_omm}
\begin{equation}
\alpha_{\mathrm{ON}}=\frac{1}{\hbar}\sum_{n}\int[d\mathbf{k}]\,\frac{\partial g(E_{n,\mathbf{k}},T)}{\partial T}\,m_{n}^{z}(\mathbf{k})\,\Omega^z_{n}(\mathbf{k}).\label{eq:ON}
\end{equation}
Since the system is two dimensional, only the out-of-plane component $m_n^z(\mathbf{k})$ enters the orbital Nernst response.

In the numerical calculations, the momentum integration is carried out in a window centered at the $K$ valley, with the cutoff chosen such that the results are converged.

\section{Quasiparticle Berry curvature and thermal transport in a fully gapped superconducting state}
\label{sec:no_BFS}

\begin{figure}[t]
\begin{centering}
\includegraphics[width=3.4in]{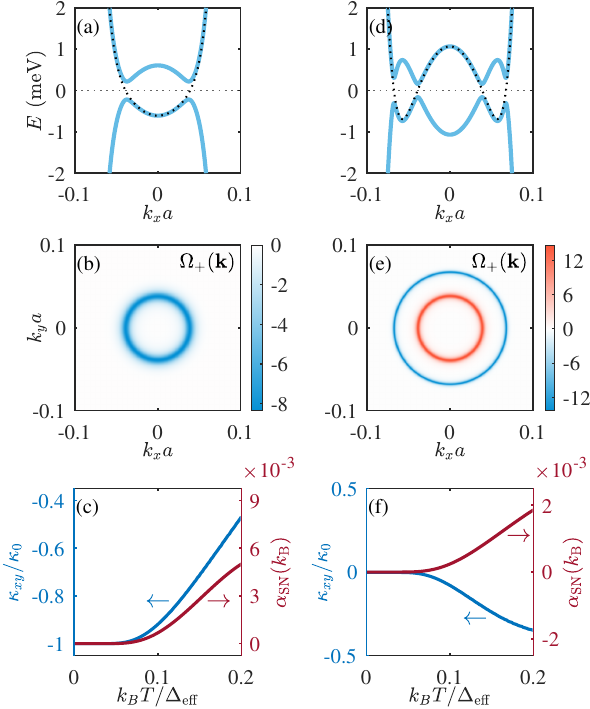}
\par\end{centering}
\caption{
(a) Quasiparticle spectrum along $k_y=0$ for $D=30~\mathrm{meV}$ and $n_e=2\times10^{11}~\mathrm{cm}^{-2}$. The blue solid curves denote the BdG quasiparticle bands, and the black dashed curve denotes the lowest-conduction-band electron dispersion measured from the chemical potential. (b) Momentum-space distribution of the Berry curvature $\Omega_+^z(\mathbf{k})$ of the upper quasiparticle band near the $K$ point for the same parameters. (c) Temperature dependence of the thermal Hall conductivity $\kappa_{xy}/\kappa_0$ and the spin Nernst coefficient $\alpha_{\mathrm{SN}}$, where $\kappa_0=\pi k_{\mathrm B}^2T/(6\hbar)$. The blue and brown curves correspond to $\kappa_{xy}/\kappa_0$ and $\alpha_{\mathrm{SN}}$, respectively, and are read from the left and right axes as indicated by the arrows. The temperature is normalized by $\Delta_{\mathrm{eff}}\equiv \min_{\mathbf{k}}[E_+(\mathbf{k})-E_-(\mathbf{k})]$, the minimum energy separation between the upper and lower BdG quasiparticle bands. (d)--(f) The corresponding quasiparticle spectrum, Berry curvature, and temperature-dependent transport coefficients for $D=60~\mathrm{meV}$ and $n_e=4\times10^{11}~\mathrm{cm}^{-2}$. The results are obtained from the self-consistent BdG Hamiltonian without trigonal warping described in Appendix~\ref{app:model_without_BFS}.
}\label{fig:BandBC_noBFS}
\end{figure}

We first consider the quasiparticle Berry curvature and thermal transport in a fully gapped chiral superconducting state of rhombohedral tetralayer graphene. The calculation is based on the self-consistent two-band model summarized in Appendix~\ref{app:model_without_BFS}, which follows Ref.~\cite{NC2025Geier_theory}. In this model, the low-energy conduction band is circularly symmetric, so that the antisymmetric dispersion $\xi_a(\mathbf{k})$ in Eq.~\eqref{eq:BdG_spectrum} vanishes. For the chiral $p$-wave gap obtained from the screened Coulomb interaction in the spin- and valley-polarized state, the quasiparticle spectrum is fully gapped for the parameters considered below. We choose two representative parameter sets from the topological and topologically trivial superconducting regions. The topological case has $D=30~{\rm meV}$ and $n_e=2\times10^{11}~{\rm cm}^{-2}$, for which the normal-state Fermi surface consists of a single circular pocket. The topologically trivial case has $D=60~{\rm meV}$ and $n_e=4\times10^{11}~{\rm cm}^{-2}$, for which the normal-state Fermi surface has an annular structure. Here, $D$ is the displacement-field-induced energy parameter, such that the potential bias between the outermost graphene layers is $2D$, and $n_e$ denotes the electron density. For these two cases, we calculate the quasiparticle spectrum, the Berry curvature, the thermal Hall conductivity, and the spin Nernst coefficient.

Figures~\ref{fig:BandBC_noBFS}(a) and \ref{fig:BandBC_noBFS}(d) show the quasiparticle spectra along $k_y=0$. The blue solid curves denote the BdG quasiparticle bands, while the black dashed curves denote the lowest-conduction-band electron dispersion measured from the chemical potential. For both parameter sets, the quasiparticle spectrum is fully gapped. The corresponding Berry curvature distributions of the upper quasiparticle band near the $K$ point are shown in Figs.~\ref{fig:BandBC_noBFS}(b) and \ref{fig:BandBC_noBFS}(e). The Berry curvature is mainly distributed around the normal-state Fermi surface. This feature can be understood from Eq.~\eqref{eq:BC}, because the Berry curvature is enhanced when the energy difference between the two BdG bands is small. For the topological single-Fermi-surface case, the Berry curvature forms a single ring in momentum space. For the topologically trivial annular-Fermi-surface case, the Berry curvature develops a two-ring structure. Thus, the momentum-space pattern of the quasiparticle Berry curvature is strongly correlated with the geometry of the underlying electron Fermi surface.

We next examine the Berry-curvature-induced thermal and spin transport responses. Figure~\ref{fig:BandBC_noBFS}(c) shows the temperature dependence of the normalized  thermal Hall conductivity $\kappa_{xy}/\kappa_0$ and the spin Nernst coefficient $\alpha_{\mathrm{SN}}$ for the topological case, while Fig.~\ref{fig:BandBC_noBFS}(f) shows the corresponding results for the topologically trivial case. The blue curves denote the thermal Hall conductivity normalized by $\kappa_0=\pi k_B^2T/(6\hbar)$, and the brown curves denote the spin Nernst coefficient. 
In the low-temperature limit of a fully gapped chiral superconductor, $\kappa_{xy}/\kappa_0$ is determined by the Chern number of the occupied BdG band. Consistently, in the topological case, $\kappa_{xy}/\kappa_0$ approaches a nonzero integer value at low temperature, reflecting the nonzero Chern number of the superconducting state. In the topologically trivial case, the low-temperature value approaches zero. At finite temperature, however, a topologically trivial state can still show a nonzero thermal Hall response because the Berry curvature is locally finite in momentum space. In the temperature range shown in Figs.~\ref{fig:BandBC_noBFS}(c) and \ref{fig:BandBC_noBFS}(f), the thermal Hall conductivity changes monotonically for both parameter sets. For both parameter sets, the spin Nernst coefficient $\alpha_{\mathrm{SN}}$ is nearly zero at low temperature and grows in magnitude as the temperature is increased. 
This behavior is mainly due to the thermal factor $\partial g(E_{n,\mathbf{k}},T)/\partial T$, which increases the weight of thermally excited quasiparticles as the temperature is raised. 
As a result, the spin Nernst coefficient increases.

We close this section by emphasizing two features of the present model. 
First, the normal-state dispersion satisfies $\xi_{\mathbf{k}}=\xi_{-\mathbf{k}}$, so the BdG spectrum remains fully gapped for the parameters considered here. 
However, recent studies suggest that superconducting rhombohedral graphene may host Bogoliubov Fermi surfaces when the realistic band structure is taken into account~\cite{PRB2025Yang_theory,PRL2026Yoon_RHG}. 
This motivates the analysis in Sec.~\ref{sec:with_BFS}, where trigonal warping is included and the intravalley dispersion becomes asymmetric. 
Second, when $\xi_{\mathbf{k}}=\xi_{-\mathbf{k}}$, Eq.~\eqref{eq:tauz_dH} shows that the two-band quasiparticle orbital magnetic moment vanishes because the relevant interband matrix element contains only the identity part. 
Therefore, the orbital magnetic moment and the orbital Nernst response are not considered in this section. 
In the next section, we show that the asymmetric dispersion generated by trigonal warping can give rise to both Bogoliubov Fermi surfaces and a finite quasiparticle orbital magnetic moment.

\section{Quasiparticle Berry curvature, orbital magnetic moment, and thermal transport in the presence of Bogoliubov Fermi surfaces}
\label{sec:with_BFS}

We now investigate the effect of Bogoliubov Fermi surfaces on the geometric properties and thermal transport of superconducting quasiparticles. We use the BdG description summarized in Appendix~\ref{app:model_with_BFS}, where the normal-state dispersion is extracted from the lowest conduction band of rhombohedral tetralayer graphene and includes trigonal warping~\cite{PRL2026Yoon_RHG}. 
The resulting asymmetric intravalley dispersion, $\xi_{\bf k}\neq \xi_{-\bf k}$, can produce Bogoliubov Fermi surfaces. 
We focus on such gapless cases and study how the Bogoliubov Fermi surfaces affect the quasiparticle Berry curvature, orbital magnetic moment, and thermal transport. 
In Appendix~\ref{app:fully_gapped_with_TW}, we also calculate the transport responses for fully gapped cases using the same BdG Hamiltonian as in this section. 
The results are qualitatively consistent with those in the last section.

\subsection{Berry curvature and orbital magnetic moment}

\begin{figure}[t]
\begin{centering}
\includegraphics[width=3.4in]{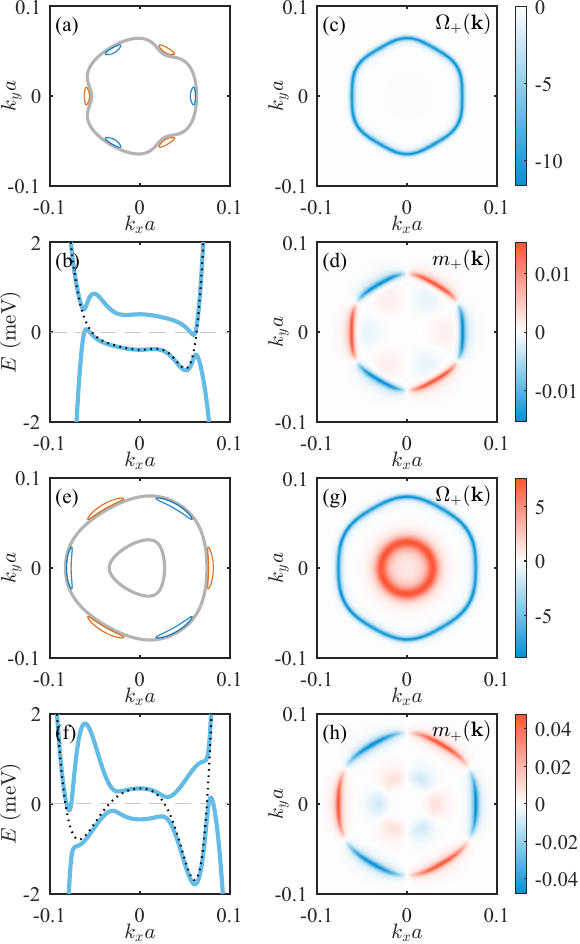}
\par\end{centering}
\caption{
(a) Normal-state (gray line) and Bogoliubov Fermi surfaces, with $U=70~\mathrm{meV}$, $n_e=5\times10^{11}~\mathrm{cm}^{-2}$, and $\Delta=4~\mathrm{meV}$. The orange and blue lines denote the Bogoliubov Fermi surfaces from the upper and lower quasiparticle bands, respectively. (b) Quasiparticle spectrum near the $K$ point for the same parameters. The blue solid curves denote the BdG quasiparticle bands, and the black dashed curve denotes the lowest-conduction-band electron dispersion measured from the chemical potential. (c) Berry curvature $\Omega^z_+(\mathbf{k})$ of the upper quasiparticle band near the $K$ point for the same parameters, in units of $\mathrm{nm}^2$. (d) Orbital magnetic moment $m^z_+(\mathbf{k})$ of the upper quasiparticle band near the $K$ point for the same parameters, in units of $\mu_B$. (e)--(h) The Fermi-surface contours, quasiparticle spectrum, Berry curvature, and orbital magnetic moment, with $U=100~\mathrm{meV}$, $n_e=7\times10^{11}~\mathrm{cm}^{-2}$, and $\Delta=8~\mathrm{meV}$. The results are obtained from the BdG Hamiltonian with trigonal warping described in Appendix~\ref{app:model_with_BFS}.
}
\label{fig:bfs_geometry}
\end{figure}

Since the Berry curvature and orbital magnetic moment directly enter the transport coefficients considered below, we first examine their momentum-space distributions. We choose two representative gapless cases with different normal-state Fermi-surface geometries. The single-Fermi-surface case has $U=70~{\rm meV}$, $n_e=5\times10^{11}~{\rm cm}^{-2}$, and $\Delta=4~{\rm meV}$, while the annular-Fermi-surface case has $U=100~{\rm meV}$, $n_e=7\times10^{11}~{\rm cm}^{-2}$, and $\Delta=8~{\rm meV}$. Here $U$ is the potential difference between the top and bottom layers of rhombohedral tetralayer graphene. These two choices allow us to compare two representative cases with different Fermi-surface geometries and different topological characters.

Figures~\ref{fig:bfs_geometry}(a) and \ref{fig:bfs_geometry}(e) show the normal-state Fermi surfaces and Bogoliubov Fermi surfaces near the $K$ point. The gray lines denote the electron Fermi surfaces, while the orange and blue lines denote the Bogoliubov Fermi surfaces from the upper and lower quasiparticle bands, respectively. In Fig.~\ref{fig:bfs_geometry}(a), the electron Fermi surface is a single trigonally warped pocket, whereas in Fig.~\ref{fig:bfs_geometry}(e) it has an annular structure with pronounced trigonal distortion. The corresponding quasiparticle spectra in Figs.~\ref{fig:bfs_geometry}(b) and \ref{fig:bfs_geometry}(f) show that the BdG branches cross zero energy at finite momenta. These zero-energy crossings form closed contours in momentum space, giving the Bogoliubov Fermi surfaces shown in Figs.~\ref{fig:bfs_geometry}(a) and \ref{fig:bfs_geometry}(e). Thus, for the parameters considered here, the chiral intravalley pairing state is not fully gapped in the presence of trigonal warping.

The momentum-space distributions of the Berry curvature and orbital magnetic moment of the upper quasiparticle band are shown in Figs.~\ref{fig:bfs_geometry}(c), (d), (g) and (h). In the single-Fermi-surface case, the Berry curvature $\Omega^z_+({\bf k})$ is mainly concentrated around a single warped ring, following the geometry of the underlying electron Fermi surface. In the annular-Fermi-surface case, the Berry curvature develops a two-ring structure, with contributions of opposite signs appearing near the inner and outer parts of the annular Fermi surface. Therefore, as in the fully gapped case discussed in Sec.~\ref{sec:no_BFS}, the Berry curvature distribution is closely related to the Fermi-surface geometry.

At the same time, the hot regions of both $\Omega^z_+({\bf k})$ and $m^z_+({\bf k})$ pass through, or lie close to, the Bogoliubov Fermi surfaces. This can be understood from the spectra in Figs.~\ref{fig:bfs_geometry}(b) and \ref{fig:bfs_geometry}(f), where the two BdG bands approach each other most closely near the Bogoliubov Fermi surfaces. According to Eqs.~(\ref{eq:BC}) and (\ref{eq:OMM}), both the Berry curvature and the orbital magnetic moment are enhanced when the energy separation between different Bogoliubov bands becomes small. However, the orbital magnetic moment is not simply proportional to the Berry curvature. Compared with $\Omega^z_+({\bf k})$, $m^z_+({\bf k})$ shows a more evident sign-changing structure, reflecting the additional matrix element involving $\tau_z\partial_{\bf k}\hat{H}^d_{\bf k}$ in Eq.~(\ref{eq:OMM}).

\subsection{Thermal Hall effect}

\begin{figure}[t]
\begin{centering}
\includegraphics[width=3.4in]{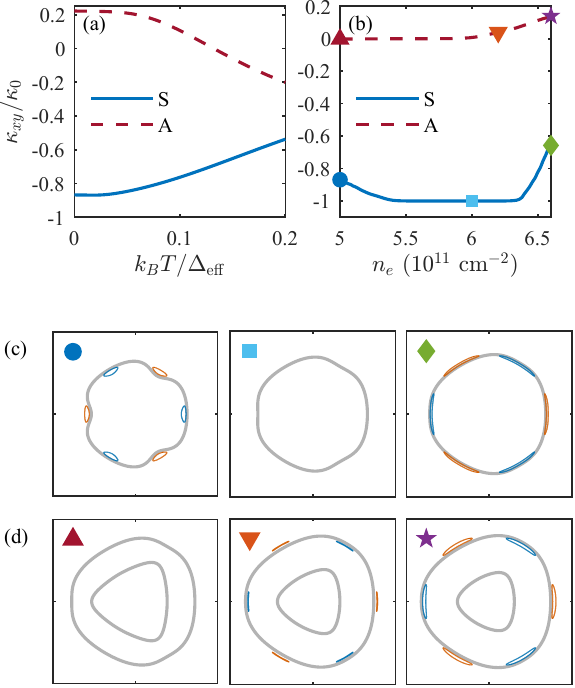}
\par\end{centering}
\caption{(a) Temperature dependence of the thermal Hall conductivity $\kappa_{xy}/\kappa_0$, with $\kappa_0=\pi k_B^2T/(6\hbar)$. The results are shown for two normal-state Fermi-surface geometries: a single Fermi surface (S) and an annular Fermi surface (A). The blue solid curve is calculated with $U=70~\mathrm{meV}$, $n_e=5\times10^{11}~\mathrm{cm}^{-2}$, and $\Delta=4~\mathrm{meV}$, while the brown dashed curve is calculated with $U=100~\mathrm{meV}$, $n_e=7\times10^{11}~\mathrm{cm}^{-2}$, and $\Delta=8~\mathrm{meV}$. The temperature is normalized by $\Delta_{\rm eff}\equiv \min_{\mathbf{k}}[E_+(\mathbf{k})-E_-(\mathbf{k})]$ for each set of parameters. (b) Electron-density dependence of $\kappa_{xy}/\kappa_0$ at $k_BT=0.015\Delta_{\rm eff}$. The blue solid curve is obtained with $U=70~\mathrm{meV}$ and $\Delta=4~\mathrm{meV}$, while the brown dashed curve is obtained with $U=100~\mathrm{meV}$ and $\Delta=8~\mathrm{meV}$. (c) and (d) Normal-state and Bogoliubov Fermi surfaces at the electron densities marked by the corresponding symbols in panel~(b). The gray lines denote the normal-state Fermi surfaces, while the orange and blue lines denote the Bogoliubov Fermi surfaces from the upper and lower quasiparticle bands, respectively. The results are obtained from the BdG Hamiltonian with trigonal warping described in Appendix~\ref{app:model_with_BFS}.}
\label{fig:thermal_hall_bfs}
\end{figure}

We next study the thermal Hall effect in the presence of Bogoliubov Fermi surfaces. Figure~\ref{fig:thermal_hall_bfs}(a) shows the temperature dependence of the normalized thermal Hall conductivity $\kappa_{xy}/\kappa_0$ for the two representative gapless cases shown in Fig.~\ref{fig:bfs_geometry}. In the zero-temperature limit, the thermal Hall response is determined by the Berry-curvature integral over the negative-energy BdG states. For a fully gapped chiral superconductor, the negative-energy states are separated from the positive-energy states by a full gap, so this integral reduces to the Chern number of the occupied BdG band. This integer Chern-number limit is no longer guaranteed when Bogoliubov Fermi surfaces are present. In the present two-band model, the Berry curvatures of the upper and lower quasiparticle bands have opposite signs. When the upper band crosses below zero energy, both BdG bands are occupied in the corresponding momentum region, and their Berry-curvature contributions partially cancel each other. When the lower band crosses above zero energy, the lower-band states in the corresponding momentum region become positive-energy states and are therefore excluded from the Berry-curvature integral over negative-energy states. In both cases, the Berry-curvature integral over the negative-energy states is changed. As a result, $\kappa_{xy}/\kappa_0$ no longer approaches a quantized value in the low-temperature limit.

In the single-Fermi-surface case, the low-temperature value of $\kappa_{xy}/\kappa_0$ is close to, but clearly different from, the quantized value as shown in Fig.~\ref{fig:BandBC_noBFS}(c) for a fully gapped topological superconductor (see also Fig.~\ref{fig:fully_gapped_with_TW}(b) in Appendix~\ref{app:fully_gapped_with_TW}). With increasing temperature, the magnitude of $\kappa_{xy}/\kappa_0$ decreases gradually. In the annular-Fermi-surface case, the low-temperature value of $\kappa_{xy}/\kappa_0$ is also nonquantized, and the coefficient further changes sign as the temperature increases. This behavior can be understood from the Berry curvature distribution in Fig.~\ref{fig:bfs_geometry}(g) and the quasiparticle spectrum in Fig.~\ref{fig:bfs_geometry}(f). At low temperature, the deviation from the reference integer value is mainly associated with the outer ring of the Berry-curvature distribution, which corresponds to quasiparticle states close to zero energy. As the temperature increases, quasiparticle states associated with the inner ring also acquire appreciable thermal weight. Since the inner and outer rings carry Berry curvatures of opposite signs, their competition leads to the sign change of $\kappa_{xy}/\kappa_0$.

To further clarify the relation between Bogoliubov Fermi surfaces and the low-temperature thermal Hall response, Fig.~\ref{fig:thermal_hall_bfs}(b) shows $\kappa_{xy}/\kappa_0$ as a function of the electron density. The calculation is performed at a low temperature, $k_BT=0.015\Delta_{\rm eff}$. We perform two density scans. The first is calculated with $U=70~{\rm meV}$ and $\Delta=4~{\rm meV}$, and the second is calculated with $U=100~{\rm meV}$ and $\Delta=8~{\rm meV}$. The corresponding normal-state and Bogoliubov Fermi surfaces at typical densities are shown in Figs.~\ref{fig:thermal_hall_bfs}(c) and \ref{fig:thermal_hall_bfs}(d).

For the density scan with single electronic Fermi surfaces, the Bogoliubov Fermi surfaces first appear, then disappear, and finally reappear as the electron density is increased, as shown in Fig.~\ref{fig:thermal_hall_bfs}(c). Correspondingly, the low-temperature value of $\kappa_{xy}/\kappa_0$ first deviates from the quantized value, then returns close to the quantized value when the Bogoliubov Fermi surfaces disappear, and deviates again when the Bogoliubov Fermi surfaces reappear. For the density scan with annular electronic Fermi surfaces, the system evolves from a regime without Bogoliubov Fermi surfaces to a regime with Bogoliubov Fermi surfaces, as shown in Fig.~\ref{fig:thermal_hall_bfs}(d). The thermal Hall coefficient follows the same trend, changing from a nearly quantized value to a nonquantized value.

These results show that the low-temperature thermal Hall response is sensitive to both the presence and the size of the Bogoliubov Fermi surfaces. This can be seen by comparing Fig.~\ref{fig:thermal_hall_bfs}(b) with the corresponding Fermi-surface contours in Figs.~\ref{fig:thermal_hall_bfs}(c) and \ref{fig:thermal_hall_bfs}(d). When the Bogoliubov Fermi surfaces are absent, $\kappa_{xy}/\kappa_0$ is close to the quantized value. When the size of the Bogoliubov Fermi surface becomes larger, the deviation from the quantized value becomes more pronounced. Therefore, the low-temperature thermal Hall conductivity provides a possible transport signature for detecting Bogoliubov Fermi surfaces in superconducting rhombohedral graphene.

\subsection{Spin and orbital Nernst effects}

\begin{figure}[h!]
\begin{centering}
\includegraphics[width=3.4in]{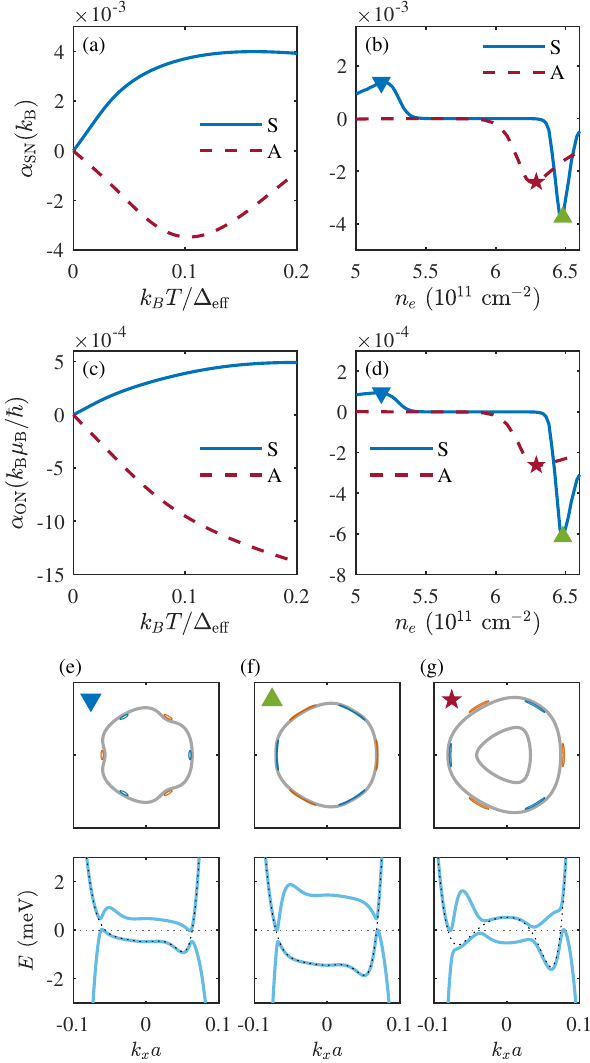}
\par\end{centering}
\caption{
(a) Temperature dependence of the spin Nernst coefficient $\alpha_{\rm SN}$. (b) Electron-density dependence of $\alpha_{\rm SN}$. (c) Temperature dependence of the orbital Nernst coefficient $\alpha_{\rm ON}$. (d) Electron-density dependence of $\alpha_{\rm ON}$. The results are shown for two normal-state Fermi-surface geometries: a single Fermi surface (S) and an annular Fermi surface (A). In panels~(a) and (c), the blue solid curve is calculated with $U=70~\mathrm{meV}$, $n_e=5\times10^{11}~\mathrm{cm}^{-2}$, and $\Delta=4~\mathrm{meV}$, while the brown dashed curve is calculated with $U=100~\mathrm{meV}$, $n_e=7\times10^{11}~\mathrm{cm}^{-2}$, and $\Delta=8~\mathrm{meV}$. In panels~(b) and (d), the blue solid curve is obtained with $U=70~\mathrm{meV}$ and $\Delta=4~\mathrm{meV}$, while the brown dashed curve is obtained with $U=100~\mathrm{meV}$ and $\Delta=8~\mathrm{meV}$. For the two density scans, the temperature is fixed at $k_BT=0.015\Delta_{\rm eff}$. (e)--(g) Normal-state and Bogoliubov Fermi surfaces and quasiparticle spectra at the electron densities marked by the corresponding symbols in panels~(b) and (d). In each panel, the upper subpanel shows the Fermi surfaces, and the lower subpanel shows the quasiparticle spectrum along $k_y=0$. The results are obtained from the BdG Hamiltonian with trigonal warping described in Appendix~\ref{app:model_with_BFS}.
}
\label{fig:nernst_bfs}
\end{figure}

We finally discuss the spin and orbital Nernst effects in the presence of Bogoliubov Fermi surfaces. Figure~\ref{fig:nernst_bfs}(a) shows the temperature dependence of the spin Nernst coefficient $\alpha_{\rm SN}$ for the two representative gapless cases shown in Fig.~\ref{fig:bfs_geometry}. Compared with the fully gapped results in Figs.~\ref{fig:BandBC_noBFS}(c) and \ref{fig:BandBC_noBFS}(f) (see also Fig.~\ref{fig:fully_gapped_with_TW}(c) in Appendix~\ref{app:fully_gapped_with_TW}), the low-temperature spin Nernst response is strongly enhanced when Bogoliubov Fermi surfaces are present. This difference can be understood from the thermal factor in Eq.~(\ref{eq:SN}). The factor $\partial g(E_{n,{\bf k}},T)/\partial T$ is strongest near zero energy, and at low temperature it is sizable only in a narrow energy window around zero energy. In the fully gapped case, there are no quasiparticle states near zero energy, so $\alpha_{\rm SN}$ remains nearly zero at low temperature. By contrast, in the present gapless case, the Bogoliubov Fermi surfaces correspond to zero-energy quasiparticle states. In both cases shown in Fig.~\ref{fig:bfs_geometry}, the Berry curvature is mainly distributed near the Bogoliubov Fermi surfaces. Therefore, the low-energy quasiparticle states selected by $\partial g/\partial T$ overlap strongly with the Berry-curvature hot regions. This overlap is the main reason why the spin Nernst coefficient is enhanced at low temperature. 

We also find that the sign of $\alpha_{\rm SN}$ in the gapless annular-Fermi-surface case is opposite to that in the fully gapped annular-Fermi-surface case shown in Fig.~\ref{fig:BandBC_noBFS}(f). 
This difference comes from the different low-energy Berry-curvature regions selected by the thermal factor. In the fully gapped case, the quasiparticle spectrum in Fig.~\ref{fig:BandBC_noBFS}(d) shows that the inner ring of the Berry-curvature distribution has lower energy, so it gives the dominant contribution to $\alpha_{\rm SN}$ at low temperature. By contrast, in the presence of Bogoliubov Fermi surfaces, Fig.~\ref{fig:bfs_geometry}(f) shows that the quasiparticle states closest to zero energy are mainly associated with the outer ring of the Berry-curvature distribution. Since the inner and outer Berry-curvature rings have opposite signs, the dominant contribution changes sign, leading to the sign reversal of $\alpha_{\rm SN}$. 
The same picture also explains the nonmonotonic temperature dependence of $\alpha_{\rm SN}$ for the annular-Fermi-surface case in Fig.~\ref{fig:nernst_bfs}(a). Although the outer Berry-curvature ring dominates in the low-temperature regime, Fig.~\ref{fig:bfs_geometry}(f) shows that the inner-ring states are also close to zero energy. As the temperature increases, these inner-ring states acquire a larger thermal weight. Their contribution has the opposite sign to that from the outer ring and therefore partially cancels it. As a result, the magnitude of $\alpha_{\rm SN}$ first increases and then decreases in the temperature range shown.

Figure~\ref{fig:nernst_bfs}(c) shows the temperature dependence of the orbital Nernst coefficient $\alpha_{\rm ON}$. This response is absent in the model with circularly symmetric normal-state dispersion of Sec.~\ref{sec:no_BFS}, because the quasiparticle orbital magnetic moment vanishes when $\xi_{\bf k}=\xi_{-\bf k}$. Since the Bogoliubov Fermi surfaces provide low-energy quasiparticles with finite Berry curvature and finite orbital magnetic moment, $\alpha_{\rm ON}$ also becomes finite already at low but finite temperature. This is in contrast to the fully gapped result in Fig.~\ref{fig:fully_gapped_with_TW}(d), where $\alpha_{\rm ON}$ remains close to zero at low temperature and grows in magnitude only as the temperature is raised. For both parameter sets, $\alpha_{\rm ON}$ changes monotonically in the temperature range shown in Fig.~\ref{fig:nernst_bfs}(c). In the annular-Fermi-surface case, the orbital Nernst coefficient does not show the same nonmonotonic behavior as the spin Nernst coefficient. The reason is that the inner ring of the orbital magnetic moment distribution gives only a weak contribution to $\alpha_{\rm ON}$. Although both the Berry curvature and the orbital magnetic moment show two-ring structures, the orbital magnetic moment on the inner ring is much weaker than that on the outer ring, as shown in Fig.~\ref{fig:bfs_geometry}(h). As a result, $\alpha_{\rm ON}$ is mainly controlled by the outer-ring Berry curvature and orbital magnetic moment, even when the temperature is increased. This suppresses the cancellation between inner- and outer-ring contributions and leads to the monotonic behavior.

In addition, the two curves in Fig.~\ref{fig:nernst_bfs}(c) have opposite signs. This sign difference is related to the momentum-space regions selected by the low-energy quasiparticles. Although the outer-ring Berry curvature and orbital magnetic moment distributions in the two cases look similar in Fig.~\ref{fig:bfs_geometry}, the Bogoliubov Fermi surfaces select different parts of the sign-changing orbital magnetic moment texture. For example, along the $k_y=0$ cut, the upper quasiparticle band in the single-Fermi-surface case crosses zero energy near the positive $k_x$ side in Fig.~\ref{fig:bfs_geometry}(b). The corresponding orbital magnetic moment in Fig.~\ref{fig:bfs_geometry}(d) is mainly negative in this region. By contrast, in the annular-Fermi-surface case, the upper quasiparticle band crosses zero energy near the negative $k_x$ side in Fig.~\ref{fig:bfs_geometry}(f), where the orbital magnetic moment in Fig.~\ref{fig:bfs_geometry}(h) is mainly positive. This example shows that the sign of $\alpha_{\rm ON}$ depends on which part of the sign-changing orbital magnetic moment texture is sampled by the zero-energy quasiparticles. Therefore, the sign of $\alpha_{\rm ON}$ is sensitive to the momentum-space location of the Bogoliubov Fermi surfaces.

We then study how the Nernst responses change with electron density. Figure~\ref{fig:nernst_bfs}(b) shows $\alpha_{\rm SN}$ as a function of $n_e$ at the fixed temperature $k_BT=0.015\Delta_{\rm eff}$. For the density scan with single electronic Fermi surfaces, $\alpha_{\rm SN}$ first has a finite value, then decreases close to zero, and finally becomes finite again as $n_e$ is increased. This follows the evolution of the Bogoliubov Fermi surfaces in Fig.~\ref{fig:thermal_hall_bfs}(c), where the Bogoliubov Fermi surfaces appear, disappear, and then reappear. For the density scan with annular electronic Fermi surfaces, $\alpha_{\rm SN}$ evolves from nearly zero to a finite value, consistent with the emergence of Bogoliubov Fermi surfaces in Fig.~\ref{fig:thermal_hall_bfs}(d). Thus, the density dependence of the spin Nernst coefficient again shows that the response is closely tied to the presence of Bogoliubov Fermi surfaces.

A further feature of Fig.~\ref{fig:nernst_bfs}(b) is that the spin Nernst coefficient has peaks near the density values marked by the symbols. To understand these peaks, we show the corresponding Fermi-surface contours and quasiparticle spectra in Figs.~\ref{fig:nernst_bfs}(e)--\ref{fig:nernst_bfs}(g). These spectra show that the peaks occur when the zero-energy line cuts through the local minimum of the upper quasiparticle band and the local maximum of the lower quasiparticle band. In this case, the Bogoliubov Fermi surfaces are still small. This situation is favorable for the spin Nernst response. The Berry curvature is mainly enhanced in the momentum regions where the two BdG bands have a small energy separation. In the present cases, these regions are located near the local minimum of the upper band or the local maximum of the lower band. When these local extrema lie near zero energy, the Berry-curvature hot regions also lie close to zero energy. Since the thermal factor $\partial g/\partial T$ is sizable only near zero energy at low temperature, the overlap between the Berry-curvature hot regions and the low-energy quasiparticle states is maximized, producing peaks in $\alpha_{\rm SN}$.

When the zero-energy line cuts more deeply into the quasiparticle bands, the Bogoliubov Fermi surfaces become larger. In this regime, the zero-energy states move away from the local extrema where the Berry curvature is strongest. The overlap between the low-energy states selected by $\partial g/\partial T$ and the Berry-curvature hot regions is then reduced, and $\alpha_{\rm SN}$ decreases again. The orbital Nernst coefficient in Fig.~\ref{fig:nernst_bfs}(d) shows a similar density dependence. This similarity arises because both $\alpha_{\rm SN}$ and $\alpha_{\rm ON}$ are controlled by the same low-energy quasiparticle structure and Berry-curvature distribution.

The density scans in Figs.~\ref{fig:nernst_bfs}(b) and \ref{fig:nernst_bfs}(d) further show that the sign of the Nernst response can change with electron density. For the density scan with single electronic Fermi surfaces, the two peaks have opposite signs. This is because changing $n_e$ modifies the quasiparticle band structure and hence changes where the Bogoliubov Fermi surfaces of each BdG band appear in momentum space, as shown by the blue and orange contours in Figs.~\ref{fig:nernst_bfs}(e) and \ref{fig:nernst_bfs}(f). The low-energy quasiparticle states of each BdG band selected by the thermal factor then come from different momentum regions. These regions may carry different values and signs of the Berry curvature, the spin expectation value, and the orbital magnetic moment. Therefore, the dominant contribution to the Nernst response can change sign as the electron density is tuned, reflecting the dependence on which BdG band forms the relevant Bogoliubov Fermi surface and where it appears in momentum space. Taken together, the low-temperature enhancement and density-dependent sign changes of the Nernst responses provide possible transport signatures of Bogoliubov Fermi surfaces in superconducting rhombohedral graphene.

\section{Conclusion}
\label{sec:conclusion}

In summary, we have studied the Berry curvature, orbital magnetic moment, and thermal transport of superconducting quasiparticles in chiral superconducting rhombohedral graphene. 
We found that the quasiparticle Berry curvature is closely related to the geometry of the underlying electron Fermi surface and is enhanced near momentum regions where the BdG bands approach each other. 
The quasiparticle orbital magnetic moment has a different structure from the Berry curvature, and shows a sign-changing texture in momentum space. 
Based on these geometric quantities, we compared the transport responses of fully gapped superconducting states and gapless states hosting Bogoliubov Fermi surfaces. 
In the fully gapped case, the low-temperature thermal Hall conductivity reflects the Chern number of the occupied BdG band. 
In the presence of Bogoliubov Fermi surfaces, the distribution of occupied quasiparticle states in momentum space is modified, so that the low-temperature thermal Hall conductivity no longer approaches an integer value normalized by $\kappa_0$. 
For the Nernst responses, the fully gapped states give nearly vanishing spin and orbital Nernst signals at low temperature, whereas the zero-energy quasiparticles on the Bogoliubov Fermi surfaces strongly enhance these responses. 
The size and momentum-space location of the Bogoliubov Fermi surfaces further affect the magnitude and sign of the transport coefficients by selecting different regions of the Berry curvature and orbital magnetic moment. 
These results show that the nonquantized low-temperature thermal Hall response and the enhanced Nernst signals provide possible transport signatures of Bogoliubov Fermi surfaces in superconducting rhombohedral graphene.

\appendix

\section{Superconducting rhombohedral tetralayer graphene model neglecting trigonal warping}
\label{app:model_without_BFS}

In this appendix, we give the explicit BdG Hamiltonian used in Sec.~\ref{sec:no_BFS}. 
This model follows Ref.~\cite{NC2025Geier_theory}, where superconductivity is studied in a spin- and valley-polarized state of rhombohedral multilayer graphene. 
The low-energy conduction band is described by a circularly symmetric dispersion, and trigonal warping is neglected.

For rhombohedral $N$-layer graphene, the normal-state dispersion is written as~\cite{NC2025Geier_theory}
\begin{equation}
    \varepsilon_{\mathbf{k}}
    =
    D\sqrt{1+(k/k_0)^{2N}}
    +
    \frac{\hbar^2 k^2}{2m},
    \label{eq:appA_dispersion}
\end{equation}
where $k=|\mathbf{k}|$. 
In this work we take $N=4$ for rhombohedral tetralayer graphene. 
The parameters $k_0$ and $m$ are treated as $D$-dependent fitting parameters for the low-energy band~\cite{NC2025Geier_theory}. 
Since Eq.~\eqref{eq:appA_dispersion} depends only on $k$, it satisfies $\varepsilon_{\mathbf{k}}=\varepsilon_{-\mathbf{k}}$. 

The pairing interaction is obtained from the screened Coulomb repulsion. 
The density-density interaction is written as
\begin{equation}
    \hat{H}_{\mathrm{int}}
    =
    \frac{1}{2}
    \sum_{\mathbf{k}_1,\mathbf{k}_2,\mathbf{q}}
    V(\mathbf{q})
    \psi_{\mathbf{k}_1+\mathbf{q}}^{\dagger}
    \psi_{\mathbf{k}_2-\mathbf{q}}^{\dagger}
    \psi_{\mathbf{k}_2}
    \psi_{\mathbf{k}_1},
    \label{eq:appA_interaction}
\end{equation}
where $\psi_{\mathbf{k}}$ is the annihilation operator of spin-polarized electrons in a single valley.
The bare interaction is taken to be the Rytova-Keldysh potential,
\begin{equation}
    V(\mathbf{q})
    =
    \frac{e^2}{2\epsilon |q|(1+r_K |q|)} ,
    \label{eq:appA_RK}
\end{equation}
where $\epsilon$ is the dielectric permittivity of the surrounding medium and $r_K$ is the Rytova-Keldysh parameter.
In this work, we use $\epsilon=5\epsilon_0$ with $\epsilon_0$ the vacuum permittivity and $r_K=3~\rm nm$.

Following Ref.~\cite{NC2025Geier_theory}, the screening is treated within the random phase approximation. 
The screened interaction potential is
\begin{equation}
    \widetilde{V}(\mathbf{q})
    =
    \frac{
    V(\mathbf{q})
    }{
    1-
    V(\mathbf{q})\chi_{e}^{0,R}(\mathbf{q},0)/e^2
    } ,
    \label{eq:appA_screened}
\end{equation}
where the noninteracting charge susceptibility is
\begin{equation}
    \chi_{e}^{0,R}(\mathbf{q},i\nu_n)
    =
    \frac{e^2}{A}
    \sum_{\mathbf{k}}
    \frac{
    f(\varepsilon_{\mathbf{k}}-\mu)
    -
    f(\varepsilon_{\mathbf{k}+\mathbf{q}}-\mu)
    }{
    \varepsilon_{\mathbf{k}}
    -
    \varepsilon_{\mathbf{k}+\mathbf{q}}
    +
    i\hbar\nu_n
    }.
    \label{eq:appA_chi0}
\end{equation}
 Here, $A$ is the area of the two-dimensional system, and
$i\nu_n$ denotes the Matsubara frequencies.

Because the dispersion in Eq.~\eqref{eq:appA_dispersion} is circularly symmetric, the screened interaction can be decomposed into angular-momentum channels,
\begin{equation}
    \widetilde{V}_{l}(k,k')
    =
    \int_{0}^{2\pi}
    d\vartheta\,
    e^{il\vartheta}
    \widetilde{V}(\vartheta,k,k'),
    \label{eq:appA_Vl}
\end{equation}
where $\vartheta$ is the angle between $\mathbf{k}$ and $\mathbf{k}'$. 
The gap function is expanded as
\begin{equation}
    \Delta_{\mathbf{k}}
    =
    \sum_{l=0}^{\infty}
    \eta_l(k)e^{il\theta},
    \label{eq:appA_gap_expand}
\end{equation}
with $(k_x,k_y)=k(\cos{\theta},\sin{\theta})$.
The finite-temperature self-consistent equation for the $l$th angular-momentum channel is
\begin{equation}
    \eta_l(k,T)
    =
    -
    \int_{0}^{\infty}
    \frac{dk'k'}{8\pi^{2}}\,
    \widetilde{V}_{l}(k,k')
    \frac{\eta_l(k',T)}{E_{\mathbf{k}'}}
    \tanh
    \frac{E_{\mathbf{k}'}}{2k_{\mathrm{B}}T},
    \label{eq:appA_gap_equation}
\end{equation}
with
\begin{equation}
    E_{\mathbf{k}}
    =
    \sqrt{
    \left(\varepsilon_{\mathbf{k}}-\mu\right)^2
    +
    |\eta_l(k,T)|^2
    } .
    \label{eq:appA_quasiparticle_energy}
\end{equation}
In this work, we consider the chiral $p$-wave channel as the dominant pairing channel.

With the convention of Eq.~\eqref{eq:generic_BdG}, this model is described by
\begin{equation}
    \hat{H}_{\mathbf{k}}
    =
    \begin{pmatrix}
        D\sqrt{1+\frac{k^{8}}{k^{8}_0}}
        +
        \dfrac{\hbar^2 k^2}{2m}
        -
        \mu
        &
        \eta_1(k,T)e^{i\theta}
        \\
        \eta_1(k,T)e^{-i\theta}
        &
        -
        D\sqrt{1+\frac{k^{8}}{k^{8}_0}}
        -
        \dfrac{\hbar^2 k^2}{2m}
        +
        \mu
    \end{pmatrix}.
    \label{eq:appA_BdG}
\end{equation}

\section{Superconducting rhombohedral tetralayer graphene model in the presence of trigonal warping}
\label{app:model_with_BFS}

In this appendix, we give the BdG Hamiltonian  used in Sec.~\ref{sec:with_BFS}. 
We follow the BdG description introduced in Ref.~\cite{PRL2026Yoon_RHG} for quarter-metal superconductivity in rhombohedral graphene. 
The two-band BdG Hamiltonian is
\begin{equation}
    \hat{H}_{\mathbf{k}}
    =
    \varepsilon_{\mathbf{k}}\tau_{+}
    -
    \varepsilon_{-\mathbf{k}}\tau_{-}
    -
    \mu\tau_z
    +
    \Delta a
    \left(
    k_x\tau_x-k_y\tau_y
    \right),
    \label{eq:appB_BdG_tau}
\end{equation}
where $\tau_{\pm}=(\tau_0\pm\tau_z)/2$, $a=0.246~{\rm nm}$ is the graphene lattice constant, and $\Delta$ is the pairing amplitude. 
With the convention of Eq.~\eqref{eq:generic_BdG}, this Hamiltonian can be written as
\begin{equation}
    \hat{H}_{\mathbf{k}}
    =
    \begin{pmatrix}
        \xi_{\mathbf{k}} & \Delta a(k_x+i k_y)\\
        \Delta a(k_x-i k_y) & -\xi_{-\mathbf{k}}
    \end{pmatrix},
    \label{eq:appB_BdG_matrix}
\end{equation}
where $\xi_{\mathbf{k}}=\varepsilon_{\mathbf{k}}-\mu$.

The electron dispersion $\varepsilon_{\mathbf{k}}$ in Eq.~\eqref{eq:appB_BdG_tau} is extracted from the lowest conduction band of the eight-band Hamiltonian of rhombohedral tetralayer graphene. 
Using the sublattice basis \{1A, 1B, 2A, 2B, 3A, 3B, 4A, 4B\}, this Hamiltonian is written as~\cite{PRL2026Yoon_RHG}
\begin{equation}
H_{\mathrm{RG}}=
\begin{pmatrix}
H_{00}+\frac{U}{2}+\delta_A & H_{01} & H_{02} & 0\\
H_{01}^{\dagger} & H_{00}+\frac{U}{6} & H_{01} & H_{02}\\
H_{02}^{\dagger} & H_{01}^{\dagger} & H_{00}-\frac{U}{6} & H_{01}\\
0 & H_{02}^{\dagger} & H_{01}^{\dagger} & H_{00}-\frac{U}{2}+\delta_B
\end{pmatrix},
\label{eq:appB_HRG}
\end{equation}
with
\begin{align}
H_{00} & =\begin{pmatrix}0 & \gamma_{0}\tilde{f}\\
\gamma_{0}\tilde{f}^{*} & 0
\end{pmatrix},H_{01}=\begin{pmatrix}\gamma_{4}\tilde{f} & \gamma_{3}\tilde{f}^{*}\\
\gamma_{1} & \gamma_{4}\tilde{f}
\end{pmatrix},H_{02}=\begin{pmatrix}0 & \gamma_{2}/2\\
0 & 0
\end{pmatrix},\nonumber \\
\delta_{A} & =\begin{pmatrix}\delta & 0\\
0 & 0
\end{pmatrix},\delta_{B}=\begin{pmatrix}0 & 0\\
0 & \delta \nonumber 
\end{pmatrix}.
\end{align}
Here $U$ is the potential difference between the top and bottom layers, $\tilde{f}$ is the graphene structure factor, $\delta=-0.066~{\rm meV}$, and the hopping parameters are $\gamma_0=3160~{\rm meV}$, $\gamma_1=445~{\rm meV}$, $\gamma_2=-18.2~{\rm meV}$, $\gamma_3=-319~{\rm meV}$, and $\gamma_4=-79~{\rm meV}$~\cite{PRL2026Yoon_RHG}. 
Near the $K$ point, $\gamma_0 \tilde{f}^*\simeq v_0\pi$, where $v_0=\sqrt{3}\gamma_0 a/(2\hbar)$ and $\pi=\hbar(k_x+i k_y)$.

\begin{figure}[t]
\begin{centering}
\includegraphics[width=3.4in]{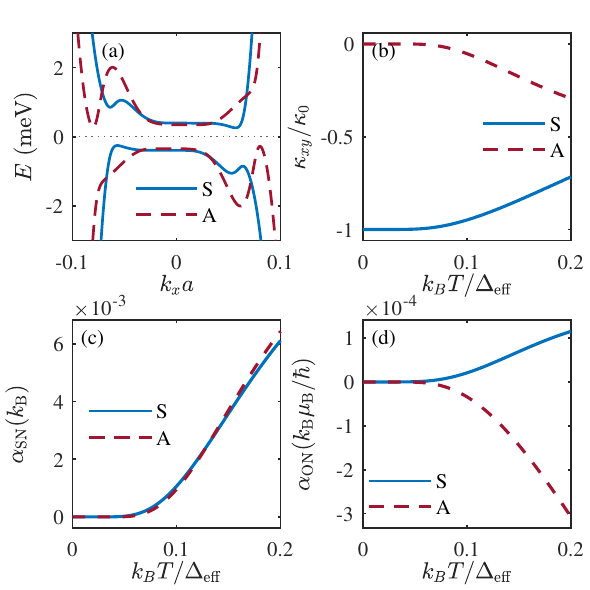}
\par\end{centering}
\caption{
(a) Quasiparticle spectra along $k_y=0$ in the fully gapped regime of the BdG Hamiltonian with trigonal warping. The results are shown for two normal-state Fermi-surface geometries: a single Fermi surface (S) and an annular Fermi surface (A). The blue solid curve is calculated with $U=70~\mathrm{meV}$, $n_e=5\times10^{11}~\mathrm{cm}^{-2}$, and $\Delta=10~\mathrm{meV}$, while the brown dashed curve is calculated with $U=100~\mathrm{meV}$, $n_e=7\times10^{11}~\mathrm{cm}^{-2}$, and $\Delta=15~\mathrm{meV}$. (b) Temperature dependence of the thermal Hall conductivity $\kappa_{xy}/\kappa_0$, with $\kappa_0=\pi k_B^2T/(6\hbar)$. (c) Temperature dependence of the spin Nernst coefficient $\alpha_{\rm SN}$. (d) Temperature dependence of the orbital Nernst coefficient $\alpha_{\rm ON}$. The temperature is normalized by $\Delta_{\rm eff}\equiv \min_{\mathbf{k}}[E_+(\mathbf{k})-E_-(\mathbf{k})]$ for each set of parameters. The results are obtained from the BdG Hamiltonian with trigonal warping described in Appendix~\ref{app:model_with_BFS}.
}
\label{fig:fully_gapped_with_TW}
\end{figure}

\section{Fully gapped results obtained from the BdG Hamiltonian with trigonal warping}
\label{app:fully_gapped_with_TW}

In this appendix, we present additional results obtained from the BdG Hamiltonian with trigonal warping described in Appendix~\ref{app:model_with_BFS}. The purpose is to provide a fully gapped reference calculation for the trigonal-warping model and compare it with the fully gapped results in Sec.~\ref{sec:no_BFS}. We use two representative parameter sets. The fully gapped single-Fermi-surface reference has $U=70~{\rm meV}$, $n_e=5\times10^{11}~{\rm cm}^{-2}$, and $\Delta=10~{\rm meV}$, while the fully gapped annular-Fermi-surface reference has $U=100~{\rm meV}$, $n_e=7\times10^{11}~{\rm cm}^{-2}$, and $\Delta=15~{\rm meV}$. Compared with the parameters used in Sec.~\ref{sec:with_BFS}, the larger pairing amplitudes remove the Bogoliubov Fermi surfaces and produce fully gapped quasiparticle spectra.

Figure~\ref{fig:fully_gapped_with_TW}(a) shows the quasiparticle spectra for these two fully gapped reference cases. The spectra are fully gapped, in contrast to the gapless spectra shown in Figs.~\ref{fig:bfs_geometry}(b) and \ref{fig:bfs_geometry}(f). Figures~\ref{fig:fully_gapped_with_TW}(b) and \ref{fig:fully_gapped_with_TW}(c) show the corresponding thermal Hall conductivity and spin Nernst coefficient. Their temperature dependences are qualitatively consistent with the fully gapped results in Sec.~\ref{sec:no_BFS}. In particular, the thermal Hall coefficient approaches the expected low-temperature integer value for the topological case and approaches zero for the topologically trivial case. The spin Nernst coefficient is close to zero at low temperature and increases as the temperature is raised. This behavior is consistent with Figs.~\ref{fig:BandBC_noBFS}(c) and \ref{fig:BandBC_noBFS}(f), and is clearly different from the low-temperature enhancement in the presence of Bogoliubov Fermi surfaces shown in Fig.~\ref{fig:nernst_bfs}(a). For the representative parameters considered here, this qualitative agreement indicates that the model with circularly symmetric normal-state dispersion used in Sec.~\ref{sec:no_BFS} captures the main transport features of the fully gapped regime.

For completeness, Fig.~\ref{fig:fully_gapped_with_TW}(d) shows the orbital Nernst coefficient. We can see that $\alpha_{\rm ON}$ remains close to zero at low temperature and grows in magnitude as the temperature is raised. This behavior contrasts with the result in Fig.~\ref{fig:nernst_bfs}(c), where the presence of Bogoliubov Fermi surfaces produces a finite orbital Nernst response already at low but finite temperature.

\bibliography{RHG}

\end{document}